\documentclass[%
 reprint,
showpacs,
nofootinbib,
 amsmath,amssymb,
 aps,
prd,
]{revtex4-1}

\usepackage{graphicx}
\usepackage{epstopdf}
\usepackage{dcolumn}
\usepackage{bm}

\begin{document}


\title{Cosmological tracking solution and the Super-Higgs mechanism}

\author{Ricardo C. G. Landim}
\email{rlandim@if.usp.br}
 \affiliation{%
 Instituto de F\'isica, Universidade de S\~ao Paulo\\
 Caixa Postal 66318,  05314-970 S\~ao Paulo, S\~ao Paulo, Brazil
}%



\date{\today}

\begin{abstract}In this paper we argue that minimal supergravity with flat K\"ahler metric and a power-law superpotential can relate the Super-Higgs mechanism for the local spontaneous supersymmetry breaking and the cosmological tracking solution, leading in turn to a late-time accelerated expansion of the universe and alleviating the coincidence problem.


\end{abstract}

\maketitle



\section{Introduction}\label{sec:level1}

Observations of Type IA Supernova indicate that the universe undergoes an accelerated expansion \cite{reiss1998, perlmutter1999}, which is dominant at present times ($\sim$ 68\%) \cite{Planck2013cosmological}.  Existing besides ordinary matter, the remaining $27\%$ of matter is an unknown form that interacts in principle only gravitationally; it is known as dark matter. The nature of the dark sector is still mysterious and it is one of the biggest challenges in the modern cosmology. The simplest dark energy candidate is the cosmological constant, whose equation of state $w_\Lambda=p_\Lambda/\rho_{\Lambda}=-1$ is in agreement with the Planck results \cite{Planck2013cosmological}. This attempt, however,  suffers from the so-called cosmological constant problem, a huge discrepancy of 120 orders of magnitude between the theoretical prediction and the observed data. Such a huge disparity motivates physicists to look into more sophisticated models. This can be done either looking for a deeper understanding of where the cosmological constant comes from, if one wants to derive it from first principles, or considering other possibilities for accelerated expansion. In the former case, an attempt is the famous KKLT model \cite{kklt2003}, and in the latter one, possibilities are even broader, with  modifications of General Relativity, additional matter fields and so on (see \cite{copeland2006dynamics, dvali2000, yin2005} and references therein).  

Among a wide range of alternatives, a scalar field is a viable candidate to be used but with a broad range of forms of the potentials. One of the first proposals was the inverse power-law potential $V(\phi)\sim \phi^{-\alpha}$, where $\alpha>0$. Although non-renormalizable, such a potential has the remarkable properties that it leads to the attractor-like behavior for the equation of state and density parameter of the  dark energy, which are nearly constants for a wide range of initial conditions, and it also alleviates the coincidence problem \cite{Zlatev:1998tr,Steinhardt:1999nw}. 

Regarding the low-energy limit of  superstring theory, supergravity is a natural option to investigate if it can furnish a model that describes the accelerated expansion of the universe, where the canonical scalar field plays the role of the dark energy. However, since supergravity with four supercharges  exists in four dimensions at most (that is, $\mathcal{N}\times 2^{D/2}=4$ for  $\mathcal{N}=1$ in $D=4$), in higher dimensions (such as $D=10$ in superstring theory) one needs more supersymmetries.  Thus minimal supergravity can be seen as an effective theory in four dimensions, and can be applied to cosmology at least as a first approximation or a toy model.\footnote{For some models of extended supergravities in cosmology, see \cite{kallosh2002supergravity,kallosh2002gauged}.}
 In the framework of minimal supergravity, Refs. \cite{Brax1999,Copeland2000} were the first attempts to describe dark energy through quintessence. Moreover,  models of holographic dark energy were embedded in minimal supergravity  in \cite{Landim:2015hqa}. 

As usual in minimal supergravity, the scalar potential can be negative, so some effort should be made in order to avoid this negative contribution. In \cite{Brax1999}  one possibility was to require $\langle W \rangle=0$, for instance.\footnote{The recent work \cite{Bergshoeff:2015tra} shows  the explicit de Sitter supergravity action, where the negative contribution of the scalar potential is avoided in another way.}

Supergravity can provide a candidate for dark matter as well, the gravitino \cite{Pagels1982}. Once local supersymmetry is broken, the gravitino acquires a mass by absorbing the goldstino, but its mass is severely constrained when considering standard cosmology. The gravitino may be the lightest superparticle (LSP) being either stable, in a scenario that preserves R-parity \cite{Pagels1982}, or unstable, but long-lived and with a  small R-parity violation \cite{Takayama2000, Buchmuller:2007ui}. Another possibility is the gravitino to be the next-to-lightest superparticle (NLSP), so that it decays into standard model particles or into  LSP.  

From the cosmological point-of-view, if the gravitino is stable its mass should be $m_{3/2}\leq 1$ keV in order not to overclose the universe \cite{Pagels1982}. Thus the gravitino may be considered as dark matter; however such value is not what is expected to solve the hierarchy problem and the gravitino describes only hot/warm dark matter; then another candidate  is needed. If the gravitino is unstable it should have $m_{3/2}\geq 10$ TeV to decay before the Big Bang Nucleosynthesis (BBN) and therefore not to conflict its results \cite{Weinberg1982}.

 These problems may be circumvented if the initial abundance of the gravitino is diluted by inflation, but since the gravitino can be produced afterward through  scattering processes when the universe  is reheated, these problems can still exist.  Furthermore, the thermal gravitino production provides an upper bound for the reheating temperature ($T_R$), in such a way that if the gravitino mass is, for instance, in the range $10$ TeV $\leq m_{3/2}\leq100$ TeV, the reheating temperature should be $T_R<10^{10}-10^{11}$ GeV \cite{Kawasaki:1994af,Kawasaki:1994bs}. Depending on which is the LSP, the gravitino mass should be even bigger than $100$ TeV \cite{nakamura2006}. On the other hand, thermal leptogenesis requires $T_R\geq 10^8-10^{10}$ GeV \cite{Davidson2002,Buchmuller2002}, which leads also to strict values for the gravitino mass. 

In this paper we expand the previous results in the literature, taking the spontaneous breaking of local supersymmetry (SUSY) into account and regarding quintessence in minimal supergravity. We consider the  scalar potential defined by the usual flat K\"aher potential $K$ and a power-law superpotential $W$,  whose scalar field $\varphi$ leads to the local SUSY breaking. After this mechanism (known also as Super-Higgs mechanism), the massive gravitino can decay into a massless scalar field $\Phi$. This scalar will play the role of the dark energy and to avoid a fifth force, the potential $V(\Phi)$ will be deduced from the original  potential $V(\varphi)$. It turns out that the potential $V(\Phi)$  corresponds to the well-known tracker behavior, whose initial conditions for the scalar field do not change the attractor solution. The advantage here is that simple choices of  $K$ and $W$ lead to both Super-Higgs mechanism and the late-time accelerated expansion of the universe in a more natural and unified way. We use natural units ($\hbar=c=1$) throughout the text.

The rest of the paper is organized in the following manner. In Sect. \ref{model} we present the local supersymmetry breaking process, assuming the flat K\"aher potential $K$ and a power-law superpotential $W$. In Sect. \ref{DE} we relate the previous scalar potential to a new one, responsible for the accelerated expansion of the universe. Sect. \ref{conclu} is reserved for conclusions.

\section{The Super-Higgs mechanism}\label{model}
 
 We start our discussion using a complex scalar field $\phi=(\phi_R+i\phi_I)/\sqrt{2}$. We set $8\pi G=M_{pl}^{-1}=1$ in the following steps for simplicity. The scalar potential in $\mathcal{N}=1$ supergravity  depends on a real function $K\equiv K(\phi^i,\phi^{i*})$, called K\"ahler potential, and a holomorphic function $W\equiv W(\phi^i)$, the superpotential. We do not consider D-terms, so the potential for one scalar field  is given by
 
  \begin{equation}V= e^{K}\left[ K_{\phi\phi^*}^{-1}\left|W_\phi+K_\phi W\right|^2-3|W|^2\right]\label{potential}
   \end{equation}

 \noindent where $K_{\phi\phi^*}\equiv \frac{\partial^2 K}{\partial \phi\partial\phi^*}$ is the K\"ahler metric, $W_\phi\equiv\frac{\partial W}{\partial \phi}$, $K_\phi\equiv\frac{\partial K}{\partial \phi}$. When supergravity is spontaneously broken, the gravitino acquires a mass given by
 
 \begin{equation}m_{3/2}=\left\langle e^{K/2}|W|\right\rangle\end{equation}
 
 \noindent where $\langle ...\rangle$  means the vacuum expectation value. 
 
\begin{figure}
\centering
\includegraphics[scale=0.55]{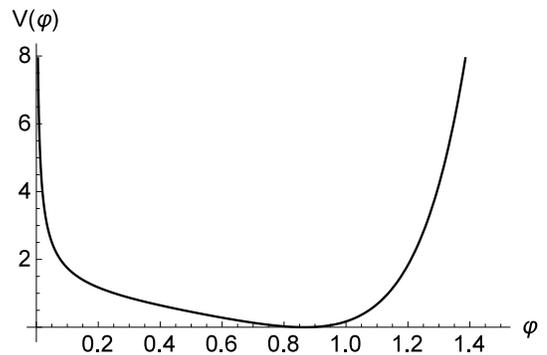}
\caption{\label{figPotCaseA} Potential $V(\varphi)$ as a function of the field $\varphi $, with $\lambda=1$ (in Planck units), for $n=3/4$.}
\end{figure}

 We use the K\"ahler potential $K_{f}=\phi\phi^*$, which leads to a flat K\"ahler metric and the superpotential  $W=\lambda^2\phi^n$, for real $n$, with $\lambda$ being a free parameter. For these choices of $K$ and $W$  the potential (\ref{potential}) is
  
 \begin{equation}
 V= \lambda^4e^{\varphi^ 2}[n^2\varphi^{2(n - 1)} + (2n - 3)\varphi^{2n} + \varphi^{2(n + 1)}]\label{potcaseA}
 \end{equation}

\noindent where  $\varphi\equiv(\sqrt{\phi_R^2+\phi_I^2})/\sqrt{2}$ is the absolute value of the complex scalar field.
We found that (\ref{potcaseA}) has extremal points at 

\begin{equation}
\varphi_1=\sqrt{1-n-\sqrt{1-n}}
 \end{equation}
\begin{equation}
\varphi_2=\sqrt{-n}
 \end{equation}
\begin{equation}\label{minPoint}
 \varphi_3=\sqrt{1-n+\sqrt{1-n}}
 \end{equation}

\noindent for $n<0$, with (\ref{minPoint}) being the global minimum, which is also  valid for the case $0<n<1$. For $n<0$ there are two local minima points ($\varphi_1$ and $\varphi_3$) which correspond to $V<0$ at these values. For negative $n$ the potential at $\varphi_3$  corresponds to the true vacuum, while the potential at $\varphi_1$ corresponds to the false vacuum. For $n>1$ $\varphi^2$ at the minimum point is negative, but this is  not allowed since $\varphi$ is  an absolute value. We consider the case of vanishing cosmological constant, so Eq. (\ref{potcaseA}) is zero at the global minimum (\ref{minPoint}) for $n=3/4$. This fractional number is the only possibility of the potential given by Eq (\ref{potcaseA}), which has $V=0$ at the minimum point. This case can be seen in Figure \ref{figPotCaseA}. Due to its steep shape, the potential cannot drive the slow-roll inflation.

 SUSY is spontaneously broken at  the minimum point $\varphi_0\equiv \varphi_3$ (\ref{minPoint}) and the gravitino becomes massive by absorbing the massless goldstino.  We estimate the gravitino mass for different values of $\lambda$, which are shown in Table \ref{tab:lambda}. The scalar mass has the same order of magnitude as the gravitino mass, therefore it may also lead to cosmological difficulties  in much the same way as the Polonyi field does \cite{Coughlan1983}. However, such a problem can be alleviated as the gravitino problem is, i.e., if the scalar mass is larger than $\mathcal{O}$(10TeV)  BBN starts after the decay of the scalar $\varphi$  has finished. From Table \ref{tab:lambda} we see that $\lambda\geq 10^{11}$ GeV does not spoil BBN.  Since the gravitino interactions are suppressed by the Planck mass, the dominant interaction is the one proportional to $M_{pl}^{-1}$ \cite{Freedman2012}, which is, for our purposes, the gravitino decay only into another complex scalar field ($\Phi$) plus its spin-1/2 superpartner ($\Psi\rightarrow \Phi+\chi$).\footnote{Although the interaction with a gauge field is also $\propto M_{pl}^{-1}$ we do not consider it for simplicity, since we have also not considered D-terms in the scalar potential. Gravitino decay into MSSM particles, such as photon and photino \cite{Kawasaki:1994af} or neutrino and sneutrino \cite{Kawasaki:1994bs}, is not considered here.} The scalar $\Phi$ is initially massless, but after the procedure we are going to follow below, it obtains a scalar potential and will be responsible for the current accelerated expansion of the universe. The decay rate for the gravitino  is $\Gamma_{3/2} \sim m_{3/2}^3/M_{pl}^2$, thus for $\lambda\geq10^{12}$ GeV, the  gravitino time decay is $\leq10^{-3}$s, hence it does not conflict BBN results. The temperature due to the decay is $T_{3/2}\sim \sqrt{\Gamma_{3/2}}\leq 10^{7}$ GeV for $\lambda\leq10^{12}$ GeV, in agreement with \cite{Kawasaki:1994af,Kawasaki:1994bs}, but with a gravitino mass one order greater than the range showed in these references. For $\lambda\geq10^{13}$ GeV we get $T_{3/2}\geq 10^{10}$ GeV.

	As a result of the gravitino decay, it also appears a massless spin-1/2 fermion $\chi$, which behaves like radiation. The $\chi$ interactions are also suppressed  by the Planck scale because they are due to four fermions terms ($\propto M_{pl}^{-2}$) or through its covariant derivative, whose interaction is with $\partial \Phi$. Since the fermion $\chi$ has a radiation behavior, its cosmological contribution is diluted as the universe expands by $a^{-4}$. When the universe is radiation-dominated, $\Omega_{\Phi}$ is small, as so $\dot{\Phi}$, which implies that the interaction due to the covariant derivative is also small. Therefore, the contribution of $\chi$ can be ignored and  the gravitino can decay into dark energy, whose associated potential for the scalar is going to be deduced from Eq. (\ref{potcaseA}). 
	
   
   \section{Dark energy} 	\label{DE}
  
 We first review the complex quintessence as a dark energy candidate, presented in \cite{Gu2001}. In this section the quintessence field will be related with the results of the last section and the  corresponding quintessence potential will be deduced from Eq. (\ref{potcaseA}). 

As seen in Eq. (\ref{potcaseA}) the scalar potential depends on the absolute value of the scalar field, as it is in fact for a complex quintessence field. As in \cite{Gu2001}, the complex quintessence can be written as $S=\Phi e^{i\theta}$, where $\Phi\equiv|S|$ is the absolute value of the scalar and $\theta$ is a phase, both depending only on time. The equations of motion for the complex scalar field in an expanding universe with FLRW metric and with a potential that depends only on the absolute value $\Phi$ (as  in our case) is \cite{Gu2001}

  \begin{equation}\label{eq:KG}
  \ddot{\Phi}+3H\dot{\Phi}+\frac{d}{d\Phi}\left(\frac{\omega^2}{2a^6}\frac{1}{\Phi^2}+V(\Phi)\right)=0,
 \end{equation}

  \noindent where $a$ is the scale factor and the first term in the brackets comes from the equation of motion for $\theta$, with $\omega$ being an integration constant interpreted as angular velocity \cite{Gu2001}. This term drives $\Phi$ away  from zero  and the factor $a^{-6}$ may make the term decrease very fast, provided that $\Phi$ does not decrease faster than $a^{-3/2}$. This situation never happens for our model, so that the contribution due to the phase component $\theta$ decreases faster than the matter density  $\rho_m$ (proportional to $a^{-3}$). Furthermore, if $\omega$ is small the complex scalar field behaves like a real scalar. From all these arguments, we neglect the complex contribution in the following.

 \begin{table}\centering
\begin{tabular}{ll}
\hline\noalign{\smallskip}
$\lambda$ (GeV)&$m_{3/2}$ (GeV)\\

\noalign{\smallskip}\hline\noalign{\smallskip}
$10^{14}$  &  $10^{10}$  \\
$10^{13}$  &  $10^{8}$  \\
$10^{12}$  &  $10^{6}$  \\
$10^{11}$  &  $10^{4}$  \\
$10^{10}$  &  $10^{2}$  \\
$10^{9}$  &  $1$  \\
$10^{8}$  &  $10^{-2}$  \\
$10^{7}$  &  $10^{-4}$  \\
$10^{6}$  &  $10^{-6}$  \\ 
  
 \noalign{\smallskip}\hline
\end{tabular}
\caption{\label{tab:lambda} Gravitino masses for different values of $\lambda$.}
\end{table}

After supergravity is spontaneously broken, the scalar field $\varphi$ may oscillate around $V(\varphi_0)=0$ and the massive gravitino can decay into the scalar field $\Phi$.  Since the  scalar $\varphi$ oscillates around $V(\varphi_0)=0$ before decay,  the following steps are naturally justified. We first shift the scalar potential from $V(\varphi)$ to $V(\varphi-\varphi_0)$, thus the minimum point goes from $\varphi_0\neq 0$ to $\varphi_0=0$.\footnote{One may wonder if this shift is valid once $\varphi$ cannot be negative, which would not be the case after this procedure. However, this shift is done in order to get a potential for the scalar $\Phi$, originated from the gravitino decay, thus the shift is justified.} The Taylor expansion of $V(\varphi)$ around the new minimum $\varphi=\varphi_0=0$ (for $n=3/4$) leads to a natural exponential term $e^{\varphi^2}\approx 1$. We finally make a change of variables $\varphi\rightarrow\Phi$. All these procedures give rise to the potential $V(\Phi)$ 


    \begin{equation}\label{potexpanded}
  V(\Phi)=\frac{M^{9/2}}{\sqrt{\Phi}} +\mathcal{O}(\Phi^{1/2}).
   \end{equation}
 
  \noindent The constant $M^{9/2}$ is written in this form for convenience. The leading-order term of this potential (\ref{potexpanded}) is an example of the well-known tracking behavior \cite{Zlatev:1998tr,Steinhardt:1999nw}. Although the field $\Phi$ is initially massless, the procedure above led to an effective potential that was absent in the original Lagrangian. We then see that such a potential has been deduced differently from that presented in \cite{Brax1999,Copeland2000}.  
	
	The parameter $\lambda$ does not have the same order of $M^{9/2}$, as long as the original shifted potential was expanded around zero. The scalar field $\varphi$ might oscillate between zero and some tiny value of the scalar potential, such as $10^{-47} $ GeV$^4$, for instance. Therefore, the order of magnitude of the  potential $V(\Phi)$ ($M^{9/2}$) is due to the magnitude of the oscillations, but its numerical value is determined phenomenologically.  The tracker condition $\Gamma\equiv VV''/V'^2=(n+1)/n>1$ (for a generic potential $V(\phi)\sim \phi^{-\alpha}$ for $\alpha>0$) is satisfied in our case, where $\alpha=1/2$. Notice that the potential (\ref{potcaseA}) can exhibit a tracking behavior only for $n<1$. 
	
	The dynamical system analysis of the quintessence field shows that the potential (\ref{potexpanded}) leads to a fixed point in the phase plane, which is a stable attractor \cite{copeland2006dynamics, Landim:2015uda}. The equation of state for the quintessence field is given by $w_\Phi=-1+\tilde{\lambda}^2/3$, where $\tilde{\lambda}\equiv -V'(\Phi)/V((\Phi)$ decreases to zero for the tracking solution. The potential Eq. (\ref{potexpanded}) gives an equation of state $w_\Phi=-0.96$, in agreement with the Planck results \cite{Planck2013cosmological}, for $\tilde{\lambda}=0.35$, which in turn is easily achieved since $\tilde{\lambda}$ decreases to zero.
 
 As is usual in quintessence models in supergravity, such as in \cite{Brax1999,Copeland2000},  different energy scales are needed to take both dynamical dark energy and local SUSY breaking into account. This  unavoidable feature is present in our model in the fact that we have needed two different free parameters $\lambda$ and $M^{9/2}$ and they are determined by phenomenological arguments.

 \section{Conclusions} \label{conclu}
 In this paper we have analyzed the scalar potential in minimal supergravity using the traditional flat K\"ahler potential and the power-law superpotential, whose  single scalar field $\varphi$  is responsible for the Super-Higgs mechanism. The gravitino decays before BBN and originates other scalar $\Phi$, which is regarded as the dark energy. Although initially massless, the scalar $\Phi$ had  its potential derived from the potential $V(\varphi)$, where the leading-order term of the expanded original potential (after a suitable change of variable) is a well-known example that satisfies the tracker condition. Therefore, the Super-Higgs mechanism and the cosmological tracking solution are not completely independent.

\begin{acknowledgments}
I thank Elcio Abdalla and Giancarlo Camilo for various suggestions and comments, during all the steps of the work. This work is supported by FAPESP Grant No. 2013/10242-1. 
\end{acknowledgments}

\bibliography{trab1}

\end{document}